\documentclass[aps, prl, twocolumn, superscriptaddress,numerical,showpacs,longbibliography,floatfix,footnoteinbib]{revtex4-2}
\usepackage{amsmath,amssymb}
\usepackage{graphicx}
\usepackage{bm}
\usepackage[usenames,dvipsnames,svgnames,table]{xcolor}
\usepackage[colorlinks=True,linkcolor=blue,citecolor=blue,urlcolor=blue]{hyperref}
\usepackage{braket}
\usepackage{bbm}
\usepackage{bbold}
\usepackage{soul}
\let\oldciteauthor=\citeauthor
\def\citeauthor#1{\hypersetup{citecolor=blue}\oldciteauthor{#1}}
\let\oldciten=\onlinecite
\def\onlinecite#1{\hypersetup{citecolor=blue}\oldciten{#1}}
\let\oldcite=\cite
\def\cite#1{\hypersetup{citecolor=blue}\oldcite{#1}}

\begin{document}
\title{Interacting Local Topological Markers: A one-particle density matrix approach for characterizing the topology of interacting and disordered states}
\author{Julia D.\ Hannukainen}\thanks{These, alphabetically listed, authors contributed equally to this work}
\author{Miguel F.\ Martínez}\thanks{These, alphabetically listed, authors contributed equally to this work}
\author{Jens H.\ Bardarson}
\author{Thomas Klein Kvorning}
\affiliation{Department of Physics, KTH Royal Institute of Technology, 106 91, Stockholm, Sweden}
\begin{abstract}
While topology is a property of a quantum state itself, most existing methods for characterizing the topology of interacting phases of matter require direct knowledge of the underlying Hamiltonian.
We offer an alternative by utilizing the one-particle density matrix formalism to extend the concept of the Chern, chiral, and Chern-Simons markers to include interactions.
The one-particle density matrix of a free-fermion state is a projector onto the occupied bands, defining a Brillouin zone bundle of the given topological class.
This is no longer the case in the interacting limit, but as long as the one-particle density matrix is gapped, its spectrum can be adiabatically flattened, connecting it to a topologically equivalent projector.
The corresponding topological markers thus characterize the topology of the interacting phase.
Importantly, the one-particle density matrix is defined in terms of a given state alone, making the local markers numerically favorable, and providing a valuable tool for characterizing topology of interacting systems when only the state itself is available.
To demonstrate the practical use of the markers we use the chiral marker to identify the topology of midspectrum eigenstates of the Ising-Majorana chain across the transition between the ergodic and many-body localized phases.
We also apply the chiral marker to random states with a known topology, and compare it with the entanglement spectrum degeneracy.
\end{abstract}
 
\maketitle
\textit{Introduction.}---Local topological markers~\cite{Kitaev20062, Prodan2010, Prodan2011, Loring2010, Bianco2011, LORING2015,Marrazzo2017,Loring2019, Hughes2019,Irsigler2019,Loring2020,Schulz-Baldes2021,Jezequel2022,Cerjan2022,d'Ornellas2022, Kvorning2022, Chen-Wei2023, franca2023} are topological invariants particularly useful for characterizing the topology of materials lacking translation symmetry.
The Chern~\cite{Bianco2011}, chiral, and Chern-Simons markers~\cite{Kvorning2022,Hannukainen2024} provide general analytic expressions for local markers for free fermion topological states protected by local symmetries, conveniently characterizing their topological phases in practice.
All three local markers are expressed in terms of the single-particle density matrix, so the topology of a given free fermion state is verified without the need of its parent Hamiltonian.
Another advantage of the single-particle density matrix formalism is that it generalizes to interacting systems through the one-particle density matrix~\cite{Penrose1956,Koch2001,Bera2015,Bera2017,Kells2018,Hopjan2020,Hopjan2021}.

Topological phases are characterized through topological equivalence under local unitary transformations, where two states are topologically equivalent if they are connected by a symmetry-preserving local unitary operator~\cite{Wen2010}.
The one-particle density matrix is relevant to those interacting topological phases, protected by local symmetries, that include a free fermion point represented by Gaussian states.
From now on we will only discuss these phases and refer to them simply as (interacting) topological phases.

Ways of characterising the topology of interacting states include computing the topological response function for interacting many body systems~\cite{Niu1985,Chen2014,Zaletel2014,Wang2015,Altland2015}, and the single particle Green's function invariants characterizing topological phases of interacting Hamiltonians that are adiabatically connected to a single particle Hamiltonian~\cite{Hughes08, Wang2010, Gurarie2011, LeiWang2012, Wang2012, Xie2012,Manmana2012,zhao2023}.
There also exist local topological markers targeting noncrystalline materials~\cite{Markov2021,franca2023}.
While these approaches necessitate knowledge of a parent Hamiltonian, topology is an inherent property of the state itself, and it is theoretically possible to formulate topological invariants in terms of the state of interest alone.
This has numerical benefits and is important for characterizing, for example, the topology of midspectrum states in many-body localized phases~\cite{Huse2013,Kjall2014,Pekker2014,Laflorencie2022}, as well as quantum engineered states~\cite{Roy.2020, Lang.2020, Sang.2021, Lavasani.2021, Fleckenstein2022, Kells.2023}.

In this work we take advantage of the one-particle density matrix formalism~\cite{Penrose1956,Koch2001,Bera2015,Bera2017,Kells2018}, which only requires knowledge of the specific state of interest, to expand the local topological markers to interacting topological phases. 
Specifically, we extend the Chern, chiral, and Chern-Simons markers, defined in Ref.~\onlinecite{Kvorning2022}, to characterize the topology of interacting topological states close to Gaussians.
The one-particle density matrix of a Gaussian state is a projector onto the occupied single-particle states, defining a vector bundle for which the topology is characterized by the local marker of the symmetry class.
By introducing interactions the one-particle density matrix is no longer a projector, since the corresponding state is no longer Gaussian.
However, as long as the gap in the one-particle density matrix spectrum remains open, it can be flattened to that of a projector, and the corresponding local marker is still well defined, extending the topological classification to include all states with a gapped one-particle density matrix spectrum.
The interacting local Chern, chiral, and Chern-Simons markers provide closed form expressions for characterizing topology in a broad range of systems ranging from amorphous matter~\cite{Grushin2020,Corbae2023} to many-body localized phases~\cite{Alet2018, Abanin2019,Sierant2024}.

To illustrate the advantages of these markers we analyze two sets of simulations.
In the first we use the chiral marker to characterize the topology of midspectrum eigenstates of the one dimensional Ising-Majorana chain across the transition from the ergodic to the many-body localized regime~\cite{Pekker2014, Kjall2014, Venderley2018, Huse2020, Sahay2021, Wahl2022, Laflorencie2022}, providing a direct measure of topology for the first time.
Secondly, to demonstrate the robustness of the markers, we analyze the local chiral marker for generic states with a known topology.
We compare these results with the entanglement spectrum degeneracy~\cite{Li2008, Pollmann2010, Fidkowski2010b}.

\textit{Classification of free-fermion states using the one-particle density matrix.}---The one-particle density matrix for a state $\ket{\Psi}$ is 
\begin{equation}
\label{eq:rho}
\varrho = \begin{pmatrix}
\tilde{\varrho}& \kappa\\
\kappa^\dagger&1-\tilde{\varrho}^*
\end{pmatrix},
\end{equation}
where the block matrices are defined by the expectation values $\tilde{\varrho}_{ij}=\bra{\Psi} c_i^\dagger c_j\ket{\Psi}$ and $\kappa_{ij}=\bra{\Psi} c_ic_j\ket{\Psi}$ of the fermionic creation, $c_i^\dagger$ and annihilation, $c_i$ operators~\cite{Bera2015,Bera2017,Kells2018}.
Diagonalizing $\varrho$ yields a single particle eigenbasis of natural orbitals $\ket{\phi_\alpha}$, where the corresponding eigenvalues $0\leq n_\alpha\leq 1$ are interpreted as occupations of the orbitals~\cite{Penrose1956,Koch2001,Bera2015,Bera2017, Lezama2017}.
In the translation invariant and noninteracting limit the one-particle density matrix is a projector onto the occupied single particle orbitals in momentum space. 
The image of the projector constitutes a vector space at each momenta, forming a vector bundle over the Brillouin zone~\cite{nakahara18}.
In the presence of unitary symmetries the one-particle density matrix becomes block-diagonal where each block defines a bundle.
The possible symmetries constrain the one-particle density matrix of each block, limiting the bundles to a specific family, one for each Altland-Zirnbauer class~\cite{zirnbauer96,altland97,kitaev09,Ryu2010}; the class to which a bundle belongs characterizes the topological phase of the corresponding state~\cite{kitaev09}.
The one-particle density matrix is no longer a projector in the presence of interactions, and the connection to vector bundles seems lost.
However, by only considering states that can be adiabatically transformed into a Gaussian state while preserving the spectral gap in the one-particle density matrix, the vector bundle determining the topology of the state remains well defined by the bundle obtained from the band flattened density matrix, $\rho=[(2\varrho -1)/|2\varrho-1|+1]/2$, of the state.
This parameter range holds significant physical relevance, as many physical states, such as many-body localized eigenstates~\cite{Serbyn2013,Ros2015,Bera2015} and ground states of Hamiltonians with weak interactions, fall within this region. 
This region does not include all states equivalent under non-Gaussian local unitary transformations---such as those reducing the $\mathbb{Z}$ classification of interacting time reversal invariant states in one dimension to $\mathbb{Z}_8$~\cite{Fidkowski2010}---rendering the number of topological equivalence classes equal to the noninteracting case.

\textit{Local topological markers in terms of the one-particle density matrix}---The $\mathbb{Z}$-invariant Chern marker~\cite{Bianco2011,Kvorning2022} in even dimensions, and the $\mathbb{Z}$-invariant chiral- and $\mathbb{Z}_2$-invariant Chern-Simons- markers~\cite{Kvorning2022} in odd dimensions are all formulated in terms of a single-particle density matrix.
Incorporating the notion of the interacting one-particle density matrix expands the use of these local markers to become a valuable tool with which to characterize topology of interacting states, both in crystalline and disordered settings.
The local chiral marker~\cite{Kvorning2022}, which characterizes topological phases in odd dimensions with a chiral constraint $S$ such that $\{\rho,S\}=S$, $S^2=1$, is defined as
\begin{equation}
\label{eq:local-chiral-winding}
\nu(\mathbf{r})= \gamma_D \varepsilon^{i_{1},\dotsc,i_{D}}\sum_\alpha[\rho SX_{i_{1}}\rho\dotsb\rho X_{i_{D}}\rho]_{(\mathbf{r}\alpha),(\mathbf{r}\alpha)},
\end{equation}
where the dimension dependent coefficient $\gamma_D=-4(8\pi i)^{(D-1)/2}[(D+1)/2]!/(D+1)!$, is derived in the supplemental material.
$X_i$ are the Cartesian position operators where the subscripts $i$ stand for the $i$th component of the position $\mathbf{r}$, $D$ is the odd spatial dimension, and $\alpha$ denotes any internal degrees of freedom.

The local markers remain well-defined away from a translation-invariant limit even though the one-particle density matrix no longer defines a vector bundle, the key being the restoration of translation invariance in the long-wavelength limit, in which the coarse grained one-particle density matrix defines a vector bundle.
In practice, in the absence of translation invariance, this means that the value of the marker varies from lattice site to lattice site, and the quantized invariant is recovered by averaging the value of the marker over a large enough volume, where the size of the required volume corresponds to the level of coarse graining.

\textit{The Ising-Majorana model.}---We use the chiral marker to explore the topology of the disordered Ising-Majorana model described by the Hamiltonian
\begin{equation}
\label{eq:H}
    H=\sum_j \left( -it_j\gamma_j \gamma_{j+1} + g\gamma_j \gamma_{j+1} \gamma_{j+2} \gamma_{j+3}\right)
\end{equation}
where $\gamma_{2j-1}=c_j + c^\dagger_j$ and $\gamma_{2j}=i(c_j - c^\dagger_j)$ are Majorana operators expressed in terms of fermion creation, $c_j^\dagger$ and annihilation, $c_j$ operators. 
The parameters $t_{j}$ are uniformly distributed in the intervals $t_{2j-1} \in [0, e^{-\delta/2}]$ and $t_{2j} \in [0, e^{\delta/2}]$, and the interaction strength $g=0.5$.
The Hamiltonian in Eq.~\eqref{eq:H} has a time-reversal symmetry, restricting the one particle density matrix $\rho$ of the eigenstates of $H$ to be real.
Together with the particle-hole constraint that $\rho$ has by construction~\footnote{The particle hole constraint is $C\rho C=1-\rho$, where $C=\sigma_x K$ and $K$ represents complex conjugation.}, these symmetries enforce a chiral constraint on $\rho$ given by $S=\sigma_x$, where $\sigma_x$ is a Pauli matrix operating within the block space given by Eq.~\eqref{eq:rho}, placing $2\rho-1$ in the symmetry class BDI.
The Ising-Majorana model can host trivial and topological, $\nu=1$, many-body localized phases depending on the values of $t_j$ and $g$~\cite{Huse2013, Pekker2014, Kjall2014,Wahl2022,Laflorencie2022}.
By adding next-nearest neighbor terms to the Hamiltonian in Eq.~\eqref{eq:H} the model can host several nontrivial phases, for example the Majorana-XYZ model with $\nu=\pm 1$ considered in detail in the supplemental material.

\begin{figure}[t]
  \includegraphics[width=1\linewidth]{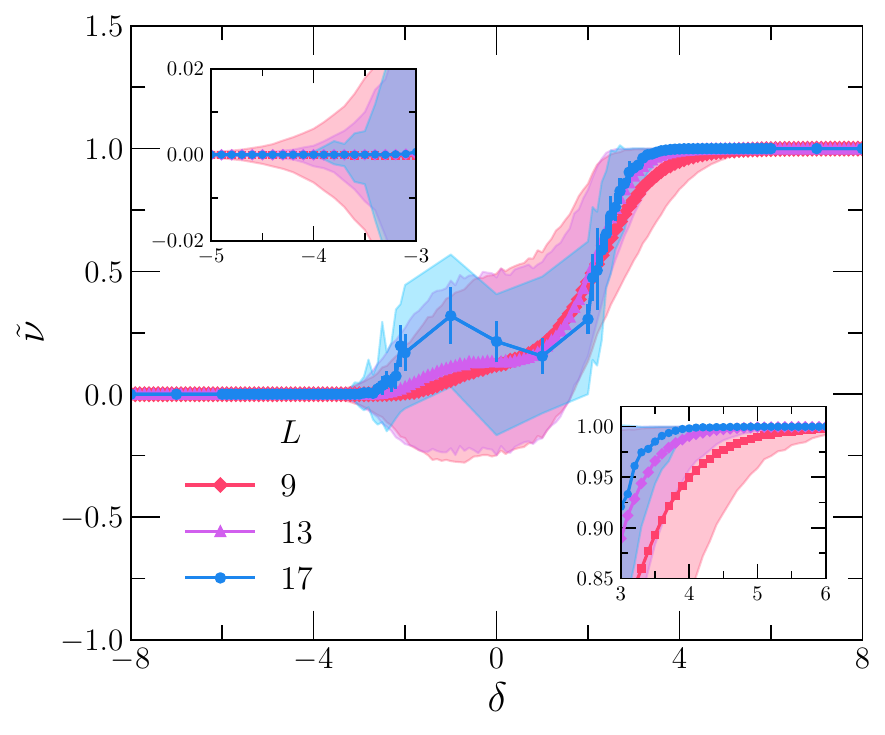}
    \caption{
    The spatially averaged chiral marker, $\nu=(1/L)\sum_{i=0}^{L-1}\nu(\bold{r}_i)$, for the Ising-Majorana model as a function of the parameter $\delta$ defining the upper bound of the disorder distribution, with interaction strength $g=0.5$, for system sizes $L=9, 13$, and $17$. 
    The plot markers represent the median $\tilde \nu$ of the ensemble of disorder realizations.
    The width of the shaded contours in the graph is defined as the smallest range of $\nu$ containing $75\%$ of the realizations.
    The insets show the median and width of the distribution for a range of $\delta$ where the shift from localized to delocalized states is predicted to occur~\protect\cite{Laflorencie2022}.
    The number of iterations for $L=9$ and $L=13$ are selected sufficiently high to ensure that the statistical errors remain imperceptible on this scale while the medians for $L=17$ are given with error bars representing a 95\% confidence interval.
     }
       
   \label{fig:Ising-Majorana1}
   \end{figure} 
   
The topology of the midspectrum eigenstates of the chain is characterized by the local chiral marker, since the one-particle density matrix is typically gapped for many-body localized states~\cite{Bera2015,Bera2017,Lezama2017}.
Fig.~\ref{fig:Ising-Majorana1} depicts the median of the spatially-averaged chiral marker $\nu$ as a function of $\delta$ for mid-spectrum energy eigenstates~\footnote{For each disorder realization we have calculated the energy eigenstate in the even fermion parity sector with energy closest to the infinite temperature value $E=0$.}.
For each state $\nu$ approaches a quantized value, $\nu=0$ for negative $\delta$ and $\nu=1$ for positive $\delta$, as $\vert\delta\vert$ increases.
For intermediate values of $\delta$ the marker is not quantized and the states are most likely not localized~\cite{Laflorencie2022}.
In this region the length scale $\xi$ setting the exponential decay of $\rho_{(\mathbf{r}\alpha),(\mathbf{r}^\prime\alpha^\prime)} \sim e^{-|\mathbf{r} - \mathbf{r}^\prime|/\xi}$ is large compared to the system-size.
The shaded region in Fig.~\ref{fig:Ising-Majorana1} contains $75\%$ of the distribution of values of $\nu$ for the different disorder samples.
This region narrows rapidly as $|\delta|$ grows, so that the topology in the localized phases is obtained from a single realization.
The insets in Fig.~\ref{fig:Ising-Majorana1} show that for large enough $|\delta|$ the distribution narrows rapidly as the system size increases, suggesting that the distribution converges to a Dirac-delta function at large system sizes.

\begin{figure*}[t]
  \includegraphics[width=1\textwidth]{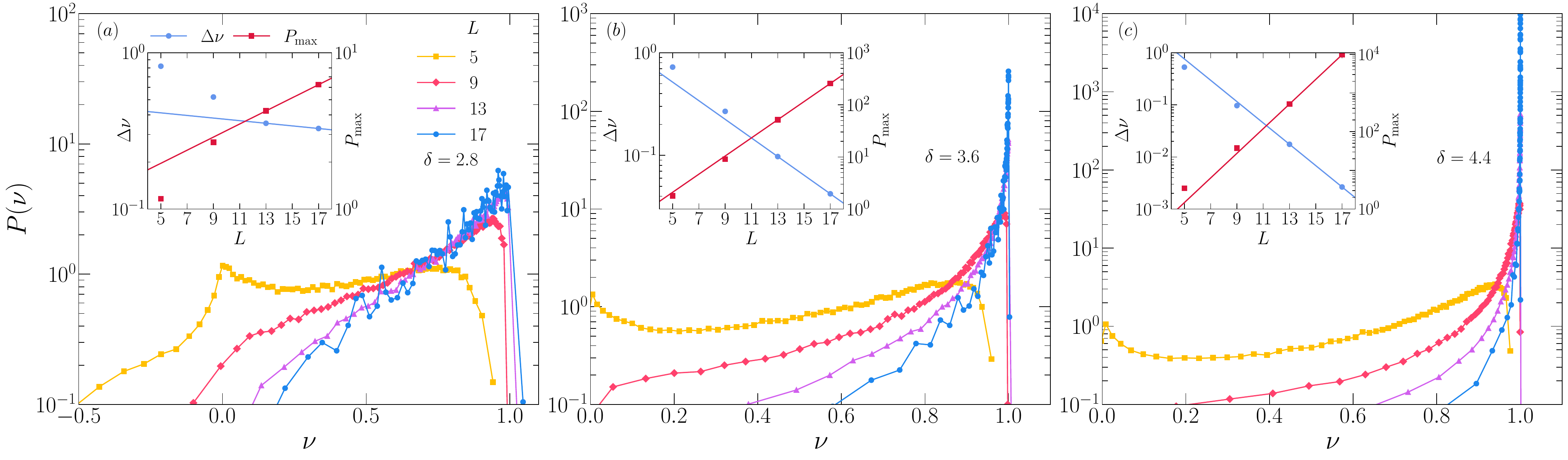}
    \caption{Numerical probability distributions of the spatially averaged chiral marker, $\nu=(1/L)\sum_{i=0}^{L-1}\nu(\bold{r}_i)$, for the Ising-Majorana model with interaction strength $g=0.5$ at (a) $\delta=2.8$, (b) $3.6$, and (c) $4.4$.
    The realizations, $N=10^5+50$ for $L<17$ and $N=2025$ for $L=17$, are partitioned into $75$ equal-size bins.
    Each $\nu$ represents the mean of the values in each bin and $P(\nu)^{-1}$ equals the number of bins (i.e., $75$) times the difference between the largest and the smallest $\nu$ value in the bin.  
    The insets show the peaks of the probability distributions $P_{\rm{max}}$ and widths $\Delta \nu$ of the intervals encompassing 75\% of the values for $\nu$, plotted against system size, and supplemented by exponential extrapolation guides based on the two largest system sizes.
    }
   \label{fig:Ising-Majorana2}
\end{figure*}

Fig.~\ref{fig:Ising-Majorana2} shows the probability density for obtaining a specific value of $\nu$ in a given disorder realization for $\delta=2.8$, $\delta=3.6$, and $\delta=4.4$.
These $\delta$ values are all in proximity to the phase transition between the ergodic and nontrivial topological phase, where an analysis using entanglement entropy becomes indeterminate for the system sizes considered here~\cite{Laflorencie2022}.
For the two larger $\delta$ values the distributions tend towards Dirac-delta functions, peaking at quantized values as the system size increases.
In the limit of infinite system size all states are therefore topological.
For $\delta=2.8$ the distribution broadens, and no conclusion can be made for the infinite system size limit.
This behavior is corroborated by the system size dependence of the peak height and width of the probability distribution, depicted in the insets of Fig.~\ref{fig:Ising-Majorana2}.
For $\delta=3.6$ and $4.4$ these are consistent with exponentials, while the data for $\delta=2.8$ is inconclusive.

\begin{figure}[t!]
  \includegraphics[width=0.95\linewidth]{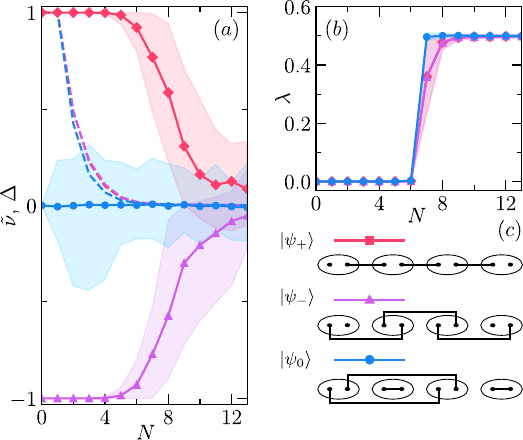}
    \caption{(a) The spatially averaged chiral marker, $\nu=(1/L)\sum_{i=0}^{L-1}\nu(\bold{r}_i)$, for the states $\ket{\psi_+}, \ket{\psi_-}$ and $\ket{\psi_0}$ as a function of the number of layers $N$ in the circuit. 
    The plot markers represent the median $\tilde \nu$ of the ensemble of disorder realizations.
    The width of the shaded contours in the graph is defined as the smallest range of $\nu$ containing $75\%$ of the realizations.
    The dashed lines show the spectral gap $\Delta$ of the one-particle density matrix.
    (b) Entanglement spectrum degeneracy as a function of the number of layers $N$. 
    (c) Schematic representation of a four-site chain for the random states at $N=0$ used in (a) and (b).
    Each circle represents one site with two Majorana operators paired by solid lines.
	}
\label{fig:ent-spec}
\end{figure}

\textit{Topology of random circuit states}---To explore the robustness of the local topological markers, and compare to the entanglement spectrum degeneracy, we apply them to random states.
In particular we analyze states as they move away from the parameter regime where the marker is expected to work.
To achieve this we transform Gaussian states with a known topology through a unitary circuit with $N$ layers, retaining the topology of the states.
The unitary gates in each layer are two-site nearest-neighbor chosen from the Haar random distribution of time-reversal- and fermion-parity-preserving gates.
We consider three classes of Gaussian states: a positive and a negative random Kitaev state $\ket{\psi_\pm}$ and a topologically trivial state $\ket{\psi_0}$.
The two topological states are defined by the two different ways the Majorana fermions couple between fermion sites $i$: $(\gamma_{2i+1} \pm \gamma_{2i}) |\psi_+\rangle = 0$, and $(\gamma_{2i-1} \pm \gamma_{2i+2}) |\psi_- \rangle = 0$, as depicted in Fig.~\ref{fig:ent-spec}(c), where the $\pm$ sign is chosen at random at each site $i$.
The trivial state is a product state with a fermion on every other site, except on the sites $(a=\lfloor L/4 \rfloor, b=L-\lfloor L/4 \rfloor)$ that are in the state $(c_a^\dagger + c_b^\dagger)\ket{0}$, with $L=26$ the number of sites.
These states satisfy the chiral constraint and are classified by the local chiral marker in Eq.~(\ref{eq:local-chiral-winding}), where $\nu= \pm 1$ for the topological states $\ket{\psi_\pm}$ and $\nu = 0$ for the trivial state.
The chiral marker, Fig.~\ref{fig:ent-spec}(a), remains quantized up to roughly five circuit layers, characterizing the topology of the three topologically distinct interacting states.
Beyond five layers the spectral gap $\Delta$ of the one-particle density matrix becomes small and the length scale $\xi$ becomes large, and the marker tends towards zero.
The reason is that for $\xi \gg L$ the marker value on each site is a random number with a random sign averaging to zero.

We compare these results with entanglement spectrum degeneracy~\cite{Li2008,Pollmann2010,Fidkowski2010b}.
The entanglement spectrum $\lbrace\varepsilon_\alpha\rbrace$ of the reduced density matrix of a region is the logarithm of its eigenvalues in an increasing order.
For a region large compared to twice the correlation length the entanglement spectrum in a topological state is degenerate~\cite{Li2008, Pollmann2010,Fidkowski2010b}. 
The degeneracy is defined by the parameter $\lambda_\alpha = (\varepsilon_\alpha- \varepsilon_{\alpha+1})/(\varepsilon_\alpha- \varepsilon_{\alpha+2})$ which is bounded by one, and where $\lambda_\alpha=0$ indicates a degeneracy in the spectrum~\footnote{If $\varepsilon_\alpha- \varepsilon_{\alpha+2} <\epsilon $, $\epsilon=10^{-14}$, we instead take the average of $\lambda_\alpha = (\varepsilon_\alpha- \varepsilon_{\alpha+3})/(\varepsilon_\alpha- \varepsilon_{\alpha+4})$. If $\varepsilon_\alpha- \varepsilon_{\alpha+4} < \epsilon$ we use $\lambda_\alpha = (\varepsilon_\alpha- \varepsilon_{\alpha+5})/(\varepsilon_\alpha- \varepsilon_{\alpha+6})$. We repeat this process until the denominator is greater than $\epsilon$.}.
Fig.~\ref{fig:ent-spec}(b) shows the half-chain entanglement spectrum degeneracy, averaged over the full entanglement spectrum, as a function of the layer number $N$ for the states obtained from $\ket{\psi_\pm}$ and $\ket{\psi_0}$.
The entanglement spectrum is degenerate for both topological states until six circuit layers, and there is no distinction between them at any $N$.
This demonstrates how states in different topological phases can share the same edge correlation and hence the same degeneracy, making it impossible to distinguish them through their entanglement spectrum.
The distribution of degeneracies for the trivial state overlaps almost everywhere with that of the topological states.
This degeneracy is caused by the Bell pair across the partition of the system, highlighting the possibility of accidental degeneracies falsely indicating that a trivial state is topological.
While this Bell pair is artificially constructed the conclusions are generic.
The marker, in contrast, captures the correct topology despite the accidental Bell pair.
It can in principle also fail but this would require a state with gapped one-particle density matrix that cannot be connected to a Gaussian state without closing the gap.
For generic states this does not happen as shown in Fig.~\ref{fig:ent-spec}(a), and we do not know how to construct such a state.
For many-body localized states this is never a problem since the $l$-bits are perturbatively connected to the Anderson orbitals~\cite{Serbyn2013,Ros2015, Bera2015}.

{\it Discussion.}---We argued that the one-particle density matrix defines a vector bundle in the presence of interactions adiabatically connected to the noninteracting limit.
This allowed us to define local topological markers that characterize the topology of nontranslationally-invariant interacting states.
An advantage of these markers is that they are defined in terms of states alone, making them a practical tool for characterizing topological phases.
To demonstrate the usefulness of the markers we characterized the topology of midspectrum states along the transition between the ergodic and many-body localized phases of the Ising-Majorana chain.
This is the first direct calculation of a topological invariant for such states, and its distribution allows a more accurate determination of the underlying phases of the Hamiltonian.

By applying the marker to random states with a known topology we verified its robustness.
The marker captures the correct topology even in cases when the entanglement spectrum degeneracy fails.
The reason is that the marker is a topological invariant while the entanglement spectrum degeneracy is only an indicator of topology.
Although there exist phases where nonlocal order parameters~\cite{pollmann2012,kvorning2020,ryu2020,wen2014} are topological invariants, they are not applicable in all instances where the local markers are.

We defined the local markers in any dimension and they are numerically efficient to calculate since they only require two-point correlation functions.
Such two-point functions are in principle measurable in experiment, which further highlights the benefit of the topological markers~\cite{Sherson2010,Ardila2018,Lukin2019,Leeuwenhoek2020,Hopjan2021B}.

{\it Acknowledgements.}---We thank D. Aceituno for helpful discussions.
The research of T.K.K. is funded by the Wenner-Gren Foundations.
This work received funding from the European Research Council (ERC) under the European Union’s Horizon 2020 research and innovation program (Grant Agreement No. 101001902), and from the Swedish Research Council (VR) through Grants No. 2019-04736 and No. 2020-00214.
J.D.H. received support from the Roland Gustafsson foundation for theoretical physics, and Stiftelsen för teknisk vetenskaplig forskning och utbildning.
The work of J.D.H. was partly performed at the Kavli Institute of Theoretical Physics.
This research was supported in part by the National Science Foundation under Grant No. NSF PHY-1748958, the Heising-Simons Foundation, and the Simons Foundation (216179, LB).
The work of J.H.B. was performed in part at the Aspen Center for Physics, which is supported by National Science Foundation grant PHY-2210452.
The computations were partially enabled by resources provided by the National Academic Infrastructure for Supercomputing in Sweden (NAISS) at the National Supercomputer Centre at Linköping University partially funded by the Swedish Research Council through grant agreement no. 2022-06725.

\bibliography{refs}

\newpage
\onecolumngrid
\setcounter{secnumdepth}{5}
\renewcommand{\theparagraph}{\bf \thesubsubsection.\arabic{paragraph}}
\renewcommand{\thefigure}{SM\arabic{figure}}
\setcounter{figure}{0} 
\appendix

\section*{Supplemental Material}

\section{The chiral marker}

In this section we derive the general form of the chiral marker in $D$ odd dimensions presented in the main text.
The chiral marker is defined as
\begin{equation}
	\nu(\bold{r})=
	2i\sum_{\alpha}\int_0^{\pi/2}\!\!\!\!\!\!\!\!d\vartheta \frac
		{%
			\varepsilon^{i_{0},\dotsc,i_{D}}%
			[P_{\vartheta}X_{i_{0}}P_{\vartheta}\dotsb X_{i_{D}}P_{\vartheta}]_{(\mathbf r,\alpha),(\mathbf r,\alpha)}%
		}%
			{[(D+1)/2]!/(2\pi i)^{(D-1)/2}},
			\label{eq:local-chiral-marker}
\end{equation}
where $S$ is the chiral constraint $\{S,\rho\}=S$ and $S^2=1$, and $X_{i_{0}}=i\partial_\vartheta$~\cite{Kvorning2022}.
The family of projectors,
\begin{equation}
P_{\vartheta}=\frac{1}{2}\big[1-\sin(\vartheta)\left(1-2\rho\right)-\cos(\vartheta)S\big]\label{eq:projector-path-chiral},
\end{equation}
interpolates between the trivial state $P_0=1/2(1-S)$ and the topological state $P_{\pi/2}=\rho$~\cite{Kvorning2022}.
The anticommutation relation $\{S,P_{\vartheta}\}=S-\cos\vartheta$ shows that the projector obeys the chiral constraint for $\vartheta=\pi/2$; the projector undergoes a topological phase transition from a trivial to a topological state obeying the chiral constraint at the point $\vartheta=\pi/2$ in the parameter space.

The form of the chiral marker in Eq.~\eqref{eq:local-chiral-marker} is simplified by using the chiral constraint and the antisymmetry of the Levi-Civita tensor.
The operator,
\begin{equation}
\begin{aligned}
\varepsilon^{i_{0},\dotsc,i_{D}}\int_0^{\pi/2}\!\!\!\!\!\!\!\!d\vartheta P_{\vartheta}X_{i_{0}}P_{\vartheta}\dotsb X_{i_{D}}P_{\vartheta}=i\varepsilon^{i_{1},\dotsc,i_{D}}\int_0^{\pi/2}\!\!\!\!\!\!\!\!d\vartheta\Big[&+ P_{\vartheta}(\partial_\vartheta P_{\vartheta})X_{i_1}P_{\vartheta}X_{i_2} P_\vartheta X_{i_3} P_\vartheta\dotsc X_{i_{D}}P_{\vartheta}\\
&-P_{\vartheta}X_{i_1}P_\vartheta(\partial_\vartheta P_\vartheta)X_{i_2} P_{\vartheta} X_{i_3}P_\vartheta\dotsb X_{i_{D}}P_{\vartheta}\\
&+P_{\vartheta}X_{i_1}P_\vartheta X_{i_2} P_\vartheta(\partial_\vartheta P_{\vartheta})X_{i_3}P_\vartheta\dotsb X_{i_{D}}P_{\vartheta}\\
&+...\\
&-P_{\vartheta}X_{i_1}P_\vartheta X_{i_2} P_\vartheta X_{i_3}P_\vartheta\dotsb (\partial_\vartheta P_{\vartheta})P_{\vartheta}\Big].
\label{eq:epsilon_expansion}
\end{aligned}
\end{equation}
only contains terms of the form $X_i\rho X_j$ are nonzero due to the commutation relation $[X_{i},S]=0$.
This means that $P_\vartheta$ can be replaced by $\rho \sin(\vartheta)$ in Eq.~\eqref{eq:epsilon_expansion}, apart from when it appears in the combination $P_\vartheta(\partial_\vartheta P_{\vartheta})$.
By expanding the derivative we have:
\begin{equation}
\begin{aligned}
P_{\vartheta} (\partial_\vartheta P_{\vartheta}) = \frac{1}{4} & \left[ -\cos(\vartheta)(1 - 2\rho) + \sin(\vartheta)S - \sin^2(\vartheta)(1 - 2\rho)S + \cos^2(\vartheta)S(1 - 2\rho) \right],
\label{eq:derivative_expansion}
\end{aligned}
\end{equation}
but, as explained in Refs.~\onlinecite{Kvorning2022,Hannukainen2024}, only the term $\frac{\rho S}{2}$ gives a nonzero contribution when inserted into Eq.~\eqref{eq:epsilon_expansion}.
Using $\{S,\rho\}=S$ to move the chiral operator to the left in each of the $D+1$ terms in Eq.~\eqref{eq:epsilon_expansion}, leads to the expression
\begin{equation}
\begin{aligned}
\varepsilon^{i_{0},\dotsc,i_{D}}\int_0^{\pi/2}\!\!\!\!\!\!\!\!d\vartheta P_{\vartheta}X_{i_{0}}P_{\vartheta}\dotsb X_{i_{D}}P_{\vartheta}=&\frac{i(D+1)}{2}\varepsilon^{i_{1},\dotsc,i_{D}}\int_0^{\pi/2}\sin^D(\vartheta) d\vartheta (\rho S X_{i_1}\rho X_{i_2} \rho  X_{i_3} \rho \dotsc X_{i_{D}}\rho)\\
=&\frac{i2^D[(\frac{D+1}{2})!]^2}{(D+1)!}\varepsilon^{i_{1},\dotsc,i_{D}} (\rho S X_{i_1}\rho X_{i_2} \rho  X_{i_3} \rho \dotsc X_{i_{D}}\rho).
\label{eq:gamma_coff_int}
\end{aligned}
\end{equation}
The chiral marker is therefore
\begin{equation}
\label{eq:appendix local-chiral-winding}
\nu(\mathbf{r})= \gamma_D \varepsilon^{i_{1},\dotsc,i_{D}}\sum_\alpha[\rho SX_{i_{1}}\rho\dotsb\rho X_{i_{D}}\rho]_{(\mathbf{r}\alpha),(\mathbf{r}\alpha)},
\end{equation}
where 
\begin{equation}
\gamma_D=-\frac{4(8\pi i)^{(D-1)/2}[(D+1)/2]!}{(D+1)!}.
\end{equation}

\section{The Majorana-XYZ model}
In this section we evaluate the topology of the XYZ-Majorana model and show that the chiral marker distinguishes between the two topological phases characterised by a winding number $\pm1$.
The XYZ-Majorana model is described by the Hamiltonian
    \begin{equation}
    \label{app:eq:H}
        H=\sum_j \left( -it_j\gamma_j \gamma_{j+1} + i t'_j \gamma_j \gamma_{j+3} + g\gamma_j \gamma_{j+1} \gamma_{j+2} \gamma_{j+3}\right).
    \end{equation}
This model hosts two different nontrivial topological phases, which in the language of Majorana edge modes corresponds to two distinct ways of connecting alternating sites on the bipartite lattice~\cite{Kitaev2001}.
In the noninteracting limit, $g=0$, the Majorana-XYZ model reduces to the Kitaev model~\cite{Kitaev2001}, which hosts two different topological phases characterized by the chiral winding number $\nu=1$ for $t_{2j-1}=0$ and $\nu=-1$ for $t'_{2j-1}=0$.
These phases survive in the presence of interactions~\cite{Fidkowski2010}.

\begin{figure}[h!]
  \includegraphics[width=0.5\linewidth]{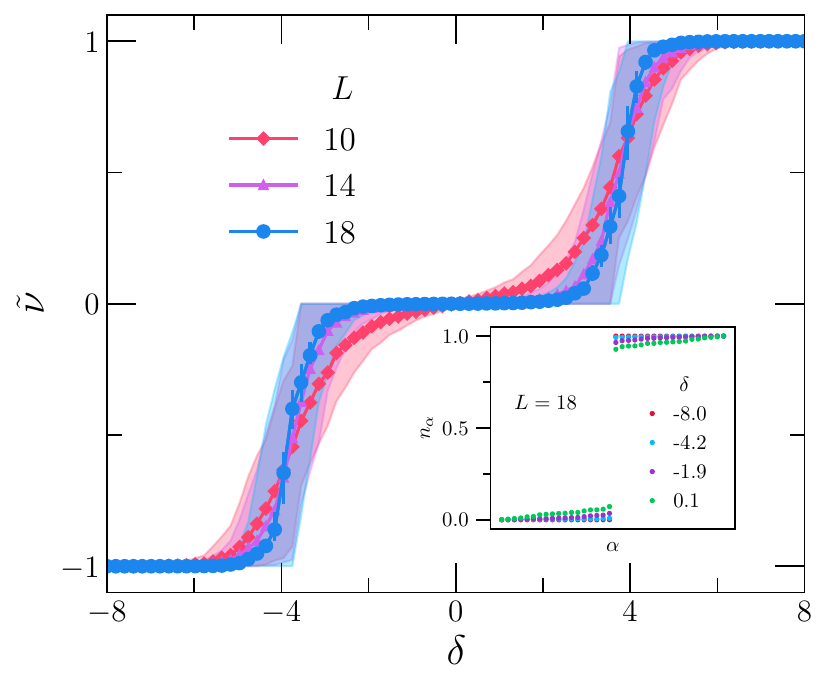}
    \caption{The spatially averaged chiral marker $\nu=(1/L)\sum_{i=0}^{L-1}\nu(\bold{r}_i)$ for the the Majorana-XYZ model as a function of the parameter $\delta$ defining the disorder strength, for interaction strength $g=10$, and system sizes $L=10, 14$, and $18$.
    The plot markers represent the median $\tilde \nu$ of the ensemble of disorder realizations while the shaded contours on the graph cover $75\%$ of all realizations. 
    These regions are identified as those having the smallest width that still retain $75\%$ of the data, for each value of $\delta$.
    The number of disorder samples is chosen so that the statistical errors for $L=10, 14$ in the median values are not visible at this scale.
    For $L=18$, the statistical errors are shown as error bars on top of the plot markers, and the data is extracted from $900$ disorder realisations.
    The inset shows the spectrum of the one particle density matrix of a single disorder realization for different values of $\delta$ and $L=18$.
	}
\label{fig:XYZ}
\end{figure} 

We use the chiral marker to characterize the topological phases of the interacting Majorana-XYZ model with added disorder, where $t_{2j-1}$ and $t'_{2j-1}$ are sampled from random uniform distributions in the intervals $t_{2j-1}\in [0, e^{\delta/2}]$ and $t'_{2j-1} \in [0, e^{-\delta/2}]$, with a fixed interaction strength $g=10$. 
These distributions are chosen such that the parameter $t_{2j-1}$ dominates when $\delta \rightarrow \infty$ and the parameter $t'_{2j-1}$ dominates when $\delta \rightarrow -\infty$, corresponding to the two different topological phases of the model in the noninteracting limit.
The spatially averaged chiral marker, $\nu=(1/L)\sum_{i}\nu(\bold{r}_i)$, is depicted in Fig.~\ref{fig:XYZ} as a function of the parameter $\delta\in[-8,8]$.
By increasing $\delta$, the phase changes from a topological phase with $\nu=-1$, to a topologically trivial phases with $\nu=0$, to another topological phase with $\nu=1$.
This result aligns with the existence of the local time-reversal-symmetry breaking unitary mapping $\gamma_j \gamma_{j+1}\leftrightarrow\gamma_j \gamma_{j+3}$ (the unitary mapping corresponds to a $\pi/2$ $z$-axis rotation in the spin language, after a Jordan-Wigner transformation $c_{i}^{\dagger}=\frac{1}{2}(X_{i}^{2}-iY_{i})\prod_{k=1}^{i-1}Z_{k}$, where $X,Y,Z$ are Pauli operators) 
This mapping requires phase boundaries to occur symmetrically around $\delta=0$. 
The shaded region in Fig.~\ref{fig:XYZ} contains $75\%$ of the distribution of values of $\nu$ for the different disorder samples.
The distribution peaks around a quantized value for increasing system sizes and in parameter regions away from the transition between topological phases.
This means that the topological phases are well defined and can typically be identified by the chiral marker for a single disorder realization, even for the relatively small system sizes considered here.
The chiral marker has the  advantage of distinguishing the two topological phases characterized by $\nu=\pm1$ without requiring phase sensitive measurements~\cite{Piet2014, Beenaker2011}.

\end{document}